\documentclass[twocolumn,article]{aastex7}
\usepackage{amsmath,amstext}
\usepackage[T1]{fontenc}
\usepackage{apjfonts} 
\usepackage{natbib}
\usepackage{longtable}
\usepackage{comment}

\newcommand{\Ha}{\hbox{{\rm H}$\alpha$}}
\newcommand{\ha}{\hbox{{\rm H}$\alpha$}}
\newcommand{\Hb}{\hbox{{\rm H}$\beta$}}
\newcommand{\hb}{\hbox{{\rm H}$\beta$}}

\newcommand{\Paa}{\hbox{{\rm Pa}$\alpha$}}

\newcommand{\Pab}{\hbox{{\rm Pa}$\beta$}}
\newcommand{\pab}{\hbox{{\rm Pa}$\beta$}}
\newcommand{\Pag}{\hbox{{\rm Pa}$\gamma$}}

\newcommand{\heii}{\hbox{[\ion{He}{2}]}}

\newcommand{\nii}{\hbox{[\ion{N}{2}]}}
\newcommand{\niiwavelen}{\hbox{[\ion{N}{2}] $\lambda6583$}}

\newcommand{\oiii}{\hbox{[\ion{O}{3}]}}

\newcommand{\oiiiwavelen}{\hbox{[\ion{O}{3}] $\lambda5007$}}

\newcommand{\nev}{\hbox{[\ion{Ne}{5}]}}

\newcommand{\siii}{\hbox{[\ion{S}{3}]}}
\newcommand{\siiiwavelen}{\hbox{[\ion{S}{3}] $\lambda9530$}}

\newcommand{\feii}{\hbox{[\ion{Fe}{2}]}}
\newcommand{\feiiwavelenone}{\hbox{[\ion{Fe}{2}] $\lambda12566$}}
\newcommand{\feiiwavelentwo}{\hbox{[\ion{Fe}{2}] $\lambda16443$}}

\usepackage{longtable}
\usepackage[utf8]{inputenc}
\usepackage{graphicx}
\usepackage{CJK}
\usepackage{makecell}

\begin{document}
\begin{CJK*}{UTF8}{gbsn}


\title{\large \bf JWST/NIRSpec Reveals a Small Population of Dominant Dust-Obscured Ionizing Sources in Galaxies at $1<z<3$}

\correspondingauthor{Nikko J. Cleri}
\email{cleri@psu.edu}

\author[0000-0001-7151-009X]{Si-rui Ge (葛思瑞)}
\affiliation{Department of Astronomy and Astrophysics, The Pennsylvania State University, University Park, PA 16802, USA}
\affiliation{School of Astronomy and Space Science, Nanjing University, Nanjing 210093, People's Republic of China}
\email{221840089@smail.nju.edu.cn}

\author[0000-0001-7151-009X]{Nikko J. Cleri}
\affiliation{Department of Astronomy and Astrophysics, The Pennsylvania State University, University Park, PA 16802, USA}
\affiliation{Institute for Computational and Data Sciences, The Pennsylvania State University, University Park, PA 16802, USA}
\affiliation{Institute for Gravitation and the Cosmos, The Pennsylvania State University, University Park, PA 16802, USA}
\email{cleri@psu.edu}

\author[0000-0001-6755-1315]{Joel Leja}
\affiliation{Department of Astronomy and Astrophysics, The Pennsylvania State University, University Park, PA 16802, USA}
\affiliation{Institute for Computational and Data Sciences, The Pennsylvania State University, University Park, PA 16802, USA}
\affiliation{Institute for Gravitation and the Cosmos, The Pennsylvania State University, University Park, PA 16802, USA}
\email{joel.leja@psu.edu}

\author[0000-0003-2536-1614]{Antonello Calabr\`o}
\affiliation{INAF Osservatorio Astronomico di Roma, Via Frascati 33, 00078 Monte Porzio Catone, Rome, Italy}
\email{antonello.calabro@inaf.it}

\author[0000-0003-0531-5450]{Vital Fern\'andez}
\affiliation{Space Telescope Science Institute, 3700 San Martin Drive, Baltimore, MD 21218, USA}
\email{vgf@stsci.edu}

\begin{abstract}
Rest-frame optical emission line diagnostics are often used to help classify ionizing sources within galaxies. However, rest-frame optical tracers can miss sources with high dust attenuation, leading to misclassification of the dominant ionizing source. Longer wavelength tracers, such as those in the near-infrared, carry the power to diagnose ionizing sources while being more robust than optical tracers to the presence of dust. The diagnostics used in this work employ the ratios of bright near-infrared emission lines \siiiwavelen, \feiiwavelenone \ and \feiiwavelentwo\ to Paschen lines (\Paa, \Pab, \Pag) in publicly-available JWST/NIRSpec MSA medium-resolution spectroscopy of 55 galaxies at $z\lesssim 3$. We compare the rest-frame near-infrared and rest-frame optical diagnostics and find that $\sim90\%$ of our sample have consistent classifications across wavelengths (49/55), while the remaining sources can be explained through ionizing radiation obscured by dust and/or elevated N/Fe abundances. We identify three objects classified as star-forming in the rest-frame optical and as active galactic nuclei (AGN) in the rest-frame near-infrared, which we interpret as obscured AGN. We also identify three objects which are classified as AGN in the rest-frame optical and star-forming in the rest-frame near-infrared. We interpret two of these objects as AGN with obscured star formation and the other with elevated N/Fe. We discuss how future spatially-resolved and/or mid-infrared spectroscopy can test the relative contributions of AGN and stars to the ionizing photon budgets of these disagreeing sources. 
\end{abstract}

\section{Introduction} \label{sec:intro}
The ratios of strong emission lines in the rest-frame optical have long been used to diagnose the primary sources of ionization in galaxies. Rest-frame optical emission line ratios can trace many of the important characteristics of active galactic nuclei (AGN) and the interstellar medium (ISM), e.g., ionization parameter, gas-phase metallicity, and shape of the ionizing continuum \citep[e.g.,][]{Baldwin_1981,Veilleux1987,Kewley2001,Kewley2002,Kewley2006,Kewley2019b,Kauffmann2003,Tremonti2004,Trump2015,Cleri2023a,Cleri2023b,Cleri2025}. These line ratios are also readily accessible from ground-based spectroscopy at low redshifts \citep[e.g.,][]{Baldwin_1981, Veilleux1987, Kauffmann2003, 2012A&A...538A...8S,Trump2015,Kewley2001,Kewley2006,Kewley2013,Kewley2019b}.

The ``BPT'' diagram\footnote{The terms ``BPT'' or ``BPT-style/BPT-like'' are often used colloquially to refer to any emission line ratio diagnostic diagram used to separate ionizing sources. For clarity, we will refer to each of the three line ratio diagnostics studied in this work explicitly; the \oiii/\hb\ versus \nii/\ha\ diagram as ``\nii-BPT'', the \siiiwavelen/\Pag \ versus \feiiwavelenone/\Pab \ diagram as ``Fe2S3-$\beta$'', and the \siiiwavelen/\Pag \ versus \feiiwavelentwo/\Paa \ diagram as ``Fe2S3-$\alpha$''.} \citep{Baldwin_1981}, perhaps the most commonly used rest-frame optical line ratio diagnostic, compares the ratios of \oiiiwavelen/\Hb \ and \niiwavelen/\Ha. These line ratios loosely trace the shape of the ionizing spectrum and ionization parameter (\oiii/\hb) and the gas-phase metallicity (\nii/\ha), and were chosen such that the lines in each ratio are close in wavelength to help mitigate the effects of dust attenuation and flux calibrations, while being readily accessible to ground-based instrumentation. 

Although the canonical rest-frame optical strong line diagnostics have been used for decades, they are calibrated for low redshifts ($z\lesssim2$) and only describe the photons from an ionizing source which can escape the galaxy \citep[e.g.,][]{Kewley2001,Kewley2006,Kewley2013,Kewley2019b,Baldwin_1981,Veilleux1987,Kauffmann2003,Trump2015,Backhaus2022,Cleri2023a,Cleri2023b,Cleri2025}. Recent studies have questioned the veracity of classifications of ionizing sources from rest-frame optical strong line ratios with increasing redshift \citep[e.g.,][]{Ubler2023,Larson2023,Cleri2025,Backhaus2025,Scholtz2025,Cameron2023a} and other studies have indicated that the fraction of heavily obscured AGN may increase as redshift increases \citep[e.g.,][]{Lambrides_2020, Buchner_2015}. 

In the presence of dust, the ionizing photon budget may not be well represented by rest-frame optical tracers. In these cases, the true ionizing photon budget may be better represented through spectroscopy of the rest-frame near-infrared (near-IR). The spectroscopic capabilities of JWST \citep{Gardner2006,Gardner_2023} with the Near-Infrared Spectrograph \citep[NIRSpec;][]{Jakobsen2022} allow for spectral coverage of the rest-frame near-IR at $z\lesssim3$. Several early works have utilized JWST/NIRSpec observations of the rest-frame near-IR to study the star-formation and dust attenuation properties of $z\lesssim3$ galaxies \citep[e.g.,][]{Reddy2023,Calabro2023,Seille2024,Neufeld2024,Cooper2025}. 

\cite{Calabro2023} leverage JWST/NIRSpec in a pilot study using spectroscopy from the Cosmic Evolution Early Release Science Survey (CEERS; \citealt{Finkelstein2025}) to test rest-frame near-IR line ratio diagnostics of ionizing sources for the effects of obscured star formation and black hole accretion. Their work introduces several diagnostics using rest-frame near-IR line ratios, including \siiiwavelen/\Pag \ versus \feiiwavelenone/\Pab \ (dubbed Fe2S3-$\beta$) and \siiiwavelen/\Pag \ versus \feiiwavelentwo/\Paa \ (dubbed Fe2S3-$\alpha$). These line ratios are convenient choices for several reasons: the Paschen lines trace the H-ionizing radiation (E > 13.6 eV) from recent star formation, while being much less sensitive to dust attenuation than UV and optical tracers \citep{Alonso-Herrero2006,Calabro2018,Kessler2020,Cleri2022,Reddy2023,Prescott2022,Gimenez-Arteaga2022,Lin2024,Seille2024,Neufeld2024}, the \siii/\Pag\ ratio traces intermediate ionization energies (23.33-34.79 eV for \siii), slightly lower than to \oiii/\hb\ in the rest-frame optical \citep[e.g.,][]{Berg2021}. \feii/Paschen-line ratios trace the gas-phase metallicity similarly to \nii/\Ha\ in the rest-frame optical, while being less susceptible to the biases of secondary N-production \citep[e.g.,][]{Calabro2023}. All ratios utilize emission lines sufficiently close in wavelength to safely neglect relative dust attenuation and instrumental calibrations. \cite{Calabro2023} prefers Fe2S3-$\beta$ as the most promising near-IR line ratio diagnostic. The \feiiwavelentwo\ line is about $15\%$ fainter than \feiiwavelenone\ but has a more limited wavelength range; both can provide useful constraints on the nature of dust-enshrouded sources as they probe longer wavelengths than the previous near-IR diagnostics \cite{Calabro2023}. These diagnostics find 5 sources which are classified as star forming in rest-frame optical diagnostics but as AGN in the Fe2S3-$\alpha$ and Fe2S3-$\beta$ rest-frame near-IR diagnostics. They interpret these sources as heavily obscured AGN, where the AGN light has been attenuated such that the host galaxy light dominates the emission line ratios in the rest-frame optical.

In this work, we expand upon the \cite{Calabro2023} Fe2S3-$\alpha$ and Fe2S3-$\beta$ diagnostics in CEERS to now use the wealth of public JWST/NIRSpec spectroscopy of $z\lesssim3$ galaxies. The paper is structured as follows: In Section \ref{sec:data}, we outline the data and sample selection. In Section \ref{sec:results}, we compare the near-IR and optical line ratio diagnostics and dust attenuation. In Section \ref{sec:discussion}, we discuss the implications of objects which disagree between the  multiwavelength diagnostics. In Section \ref{sec:summary and conclusions} we summarize the findings of this work. 
 
\section{Data}\label{sec:data}
\begin{figure*}
    \centering
    \includegraphics[width=1.0\textwidth]{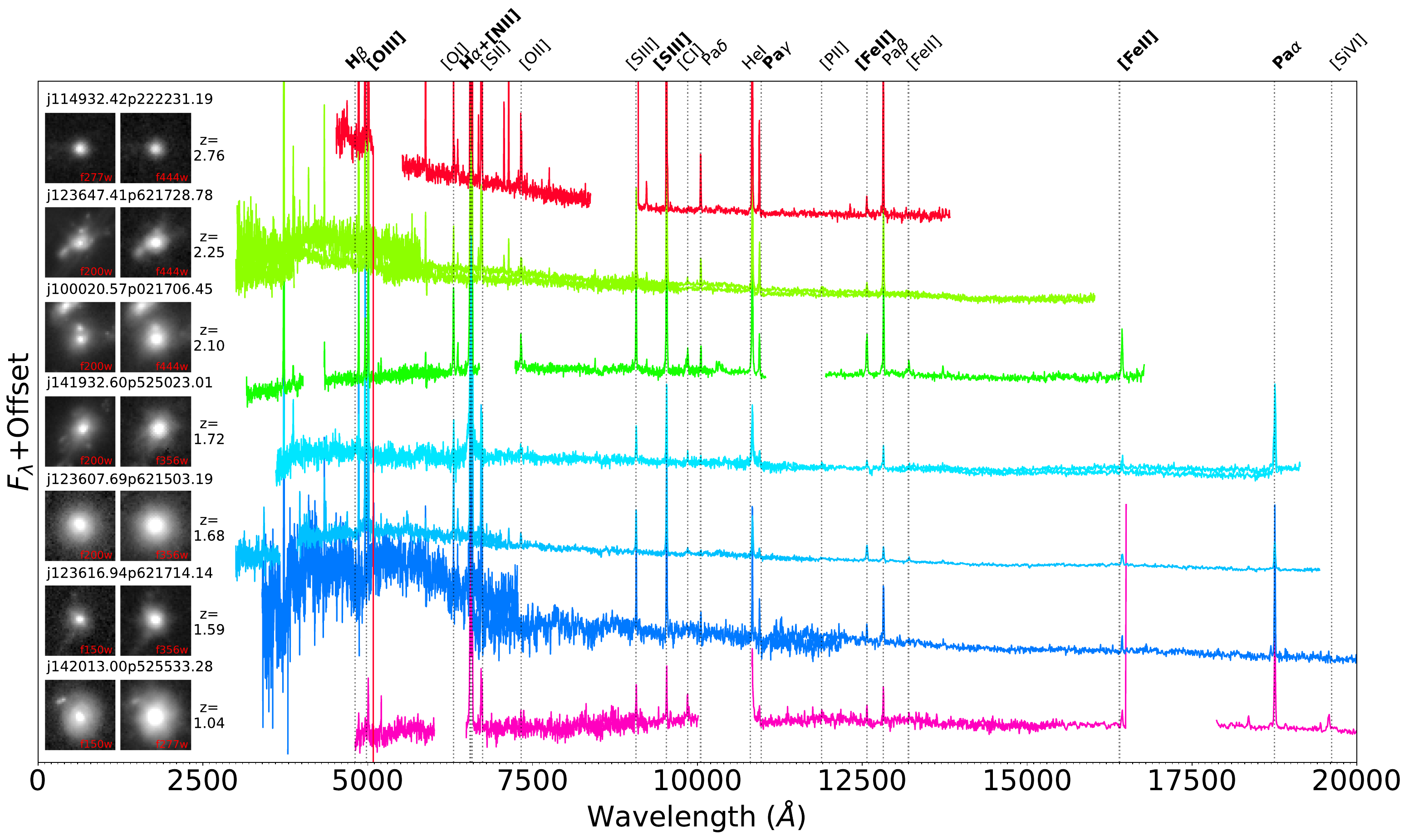}
    \caption{Rest-frame one-dimensional spectra of 7 galaxies in our sample with an offset for visualization. Each spectrum has its rest-frame optical and near-IR images on the left and its DJA ID\footnote{In the DJA spectroscopic tables, this ID is denoted as `jname': \href{https://s3.amazonaws.com/msaexp-nirspec/extractions/nirspec_public_v4.4.html}{https://s3.amazonaws.com/msaexp-nirspec/extractions/nirspec\_public\_v4.4.html}} and redshift located above and to the right of the images. We mark the locations of several prominent emission lines and note those used in the following figures in bold.}
    \label{fig: spectra}
\end{figure*}
The reduced spectra used in this work come from the DAWN JWST Archive\footnote{\href{https://dawn-cph.github.io/dja/}{https://dawn-cph.github.io/dja/}} \citep[DJA;][]{Heintz2024,deGraaff2024} The details of the NIRCam data reduction for the images are presented in \cite{Valentino_2023}, and are reduced with the grizli\footnote{\href{https://doi.org/10.5281/zenodo.8370018}{grizli: https://doi.org/10.5281/zenodo.8370018}} pipeline. NIRSpec spectroscopic data products were reduced with msaexp\footnote{\href{https://doi.org/10.5281/zenodo.7299500}{msaexp: https://doi.org/10.5281/zenodo.7299500}}. We use spectra from the JWST Advanced Deep Extragalactic Survey (JADES; ID: 1181, 1210, 1286, PIs: D. Eisenstein, N. Leutzgendorf; \citealt{Eisenstein2023a,Eisenstein2023b}), ID:1199 (PI: M. Stiavelli; \citealt{Stiavelli2023}), Systematic Mid-Infrared Instrument (MIRI) Legacy Extragalactic Survey (SMILES; ID: 1207, PI: G. Rieke; \citealt{Rieke2024,Zhu2025}), Cosmic Evolution Early Release Science (CEERS; ID: 1345, PI: S. Finkelstein; \citealt{Finkelstein2025}), Blue Jay (ID: 1810, PI: S. Belli; \citealt{Belli2024}), Assembly of Ultradeep Rest-optical Observations Revealing Astrophysics (AURORA; ID: 1914, PI: A. Shapley; \citealt{Shapley2025,Shapley2025b}), Chemical Evolution Constrained Using Ionized Lines in Interstellar Aurorae (CECILIA; ID: 2593, PI: A. Strom; \citealt{Strom2023}), and Red Unknowns: Bright Infrared Extragalactic Survey (RUBIES; ID: 4233, PIs: A. de Graaff, G. Brammer; \citealt{deGraaff2025}).

We use spectra from the G140M/F070LP, G140M/F100LP, G235M/F100LP and G395M/F290LP grating/filter pairs with $R\equiv\lambda/\Delta\lambda\sim1000$ to ensure the deblending of important emission lines, such as \nii \ and \Ha. We require spectra with robust spectroscopic redshifts (DJA grade = 3) with coverage of the emission lines in \nii-BPT and at least one of Fe2S3-$\beta$ or Fe2S3-$\alpha$. This selection yields a parent sample of 6438 spectra with robust redshifts in the range ($0<z<3.2$). Figure \ref{fig: spectra} shows the rest-frame one-dimensional spectra of 7 galaxies chosen from our sample and their rest-frame optical and near-IR images.

\begin{figure}
    \centering
    \includegraphics[width=1.0\linewidth]{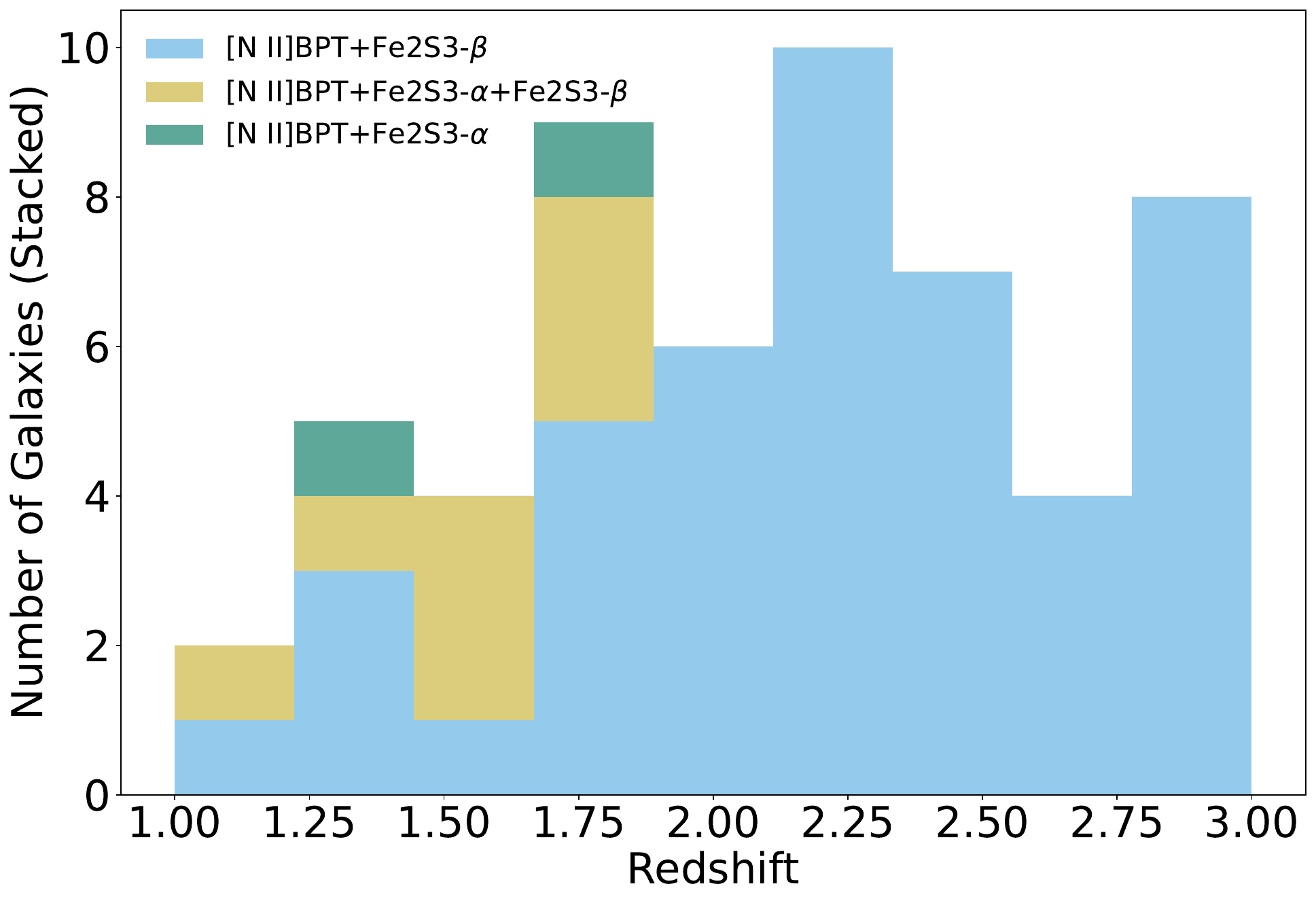}
    \caption{Stacked redshift distribution of galaxies in our sample. Blue represents galaxies with the \nii-BPT (\oiiiwavelen, \Hb, \niiwavelen \ and \Ha) and Fe2S3-$\beta$ lines (\siiiwavelen, \Pag, \feiiwavelenone \ and \Pab). Green indicates galaxies with the \nii-BPT and Fe2S3-$\alpha$ lines (\siiiwavelen, \Pag, \feiiwavelentwo \ and \Paa). Yellow indicates galaxies with all the lines from all three diagnostics detected. The details of our sample selection are outlined in Section \ref{sec:data:fluxes}.}
    \label{fig: redshift}
\end{figure}

After fitting the emission lines (described in Section \ref{sec:data:fluxes}), we make our sample of objects with \nii-BPT, Fe2S3-$\alpha$ and/or Fe2S3-$\beta$ coverage with S/N > 2 for each respective emission line, following the selection threshold of \citealt{Calabro2023}. Figure \ref{fig: redshift} shows the stacked redshift distribution of our sample of 55 total galaxies after our selection. Blue shows galaxies with the \nii-BPT and Fe2S3-$\beta$ lines detected (45), green shows galaxies with the \nii-BPT and Fe2S3-$\alpha$ lines detected (2), yellow shows galaxies with all the three diagnostics lines detected (8).

\subsection{Emission Line Fluxes}\label{sec:data:fluxes}




We use LiMe\footnote{\href{https://lime-stable.readthedocs.io/en/latest/}{https://lime-stable.readthedocs.io/en/latest/}}
 v1.3.0 \citep{Fernandez2024} to measure the emission fluxes for \oiiiwavelen, \Hb, \niiwavelen, \Ha, \siiiwavelen, \Pag, \feiiwavelenone, \Pab, \feiiwavelentwo, and \Paa. To confirm the presence of lines, LiMe employs an iterative nth-order polynomial to fit the continuum, followed by intensity thresholding that accounts for the calibration flux uncertainty. Finally, LiMe uses LmFIT \citep{Newville2014} to manage the profile constraints and fitting. In this analysis, we manually adjust, as necessary, the adjacent continuum bands to properly fit the line continuum level and avoid spectroscopic artifacts.

We fit single-component Gaussians to the \Hb, \oiii, \nii, \siii, \feii\ lines, and Paschen lines. We fit \Ha\ and the two \nii\ lines as blended features, fixing the amplitudes of \nii\ $\lambda 6584$/\nii\ $\lambda 6548$ to 2.94:1 \citep[e.g.,][]{Fischer2004} and aligning the kinematics between both lines. We also fit a two-component narrow and broad line complex for \Ha\ and \Hb. The broad component is fit with a full width at half maximum (FWHM) restricted to rest-frame $>1000$ km/s, with the line center left free. We limit the widths of \nii\ $\lambda 6584$ and \nii\ $\lambda 6548$ to that of the narrow \Ha\ component given they trace similar ionization energies.

We additionally fit individual Gaussians to \nev\ $\lambda 3427$ and \heii\ $\lambda 4687$, which we use as tracers of very-high-ionization emission \citep[>54.42 eV;][]{Berg2021,Olivier2022}, which is often attributed to the presence of an accreting massive black hole \citep[e.g.,][]{Izotov2012,Negus2023,Cleri2023a,Cleri2023b,Cleri2025}. We indicate galaxies with detections of these emission lines in Table \ref{tab:sample} in Appendix \ref{appendix:sample}.

The wavelength coverage of the NIRSpec medium gratings is 0.97$\mu$m to 5.27$\mu$m \citep{Jakobsen2022}, which sets the redshift range where all of the Fe2S3-$\alpha$ and Fe2S3-$\beta$ lines are accessible in the combined NIRSpec medium gratings at $1.00<z<1.81$ and $1.00<z<3.19$, respectively. The line ratios are composed of lines close enough in wavelength such that wavelength-dependent flux calibrations are negligible. We do not enforce the same slit position for each object between different observations, and discuss the potential implications of this choice in Section \ref{sec:discussion}.

\subsection{Ancillary Multiwavelength Data}\label{sec:data:ancillary}
We also use X-ray data from the Chandra Deep Fields, which include the deepest X-ray imaging from Chandra available anywhere on the sky (7 Ms in CDF-S and 2 Ms in CDF-N; see \citealt{Luo_2017} and \citealt{Xue_2016}, respectively). We use the $0.5-7.0$ keV luminosity selection from \cite{Xue_2011} where an object with  $L_{0.5-7 \mathrm{keV}} >3\times10^{42}~\mathrm{erg~s}^{-1}$ is classified as an AGN. 34 sources in our sample have X-ray coverage in the CDF-S (11) and CDF-N (23). Of the sources with Fe2S3-$\alpha$ and/or Fe2S3-$\beta$ detections, four have X-ray detections; of these, two are classified as X-ray AGN. 

In addition to the X-ray data, we use mid-IR photometry to find candidate obscured AGN in our sample. We use Spitzer/IRAC \citep{Werner2004,Fazio2004} photometry from the CANDELS GOODS-N multi-wavelength catalog \citep{Barro2019}. An object is classified as an unambiguous mid-IR AGN if it meets all of the following criteria from \cite{Donley2012}:
\begin{enumerate}
    \item objects are detected in all four IRAC channels (peak wavelengths 3.6, 4.6, 5.8 and 8.0 $\mu m$)
    \item $x \geq 0.08$ and $y \geq 0.15$
    \item $y \geq 1.21x - 0.27$
    \item $y \leq 1.21x + 0.27$
    \item $f_{8.0\mu m}$ > $f_{5.8\mu m}$ > $f_{4.5\mu m}$ > $f_{3.6\mu m}$
\end{enumerate}
where
\begin{align*}
    x &= \log_{10}{\frac{f_{5.8\mu m}}{f_{3.6\mu m}}}\\
    y &= \log_{10}{\frac{f_{8.0\mu m}}{f_{4.5\mu m}}}
\end{align*}

We find 18 objects with mid-IR detections (out of 23 total in the fields with IRAC coverage), two of which are classified as mid-IR AGN.

\section{Results}\label{sec:results}

In this Section, we compare the rest-frame near-IR diagnostics to the \nii-BPT diagram and analyze nebular dust attenuation indicators for our sample.

\subsection{Emission Line Ratio Diagnostics of Ionizing Sources}\label{sec:results:line_ratios}
\begin{figure*}
    \centering
    \includegraphics[width=1.0\textwidth]{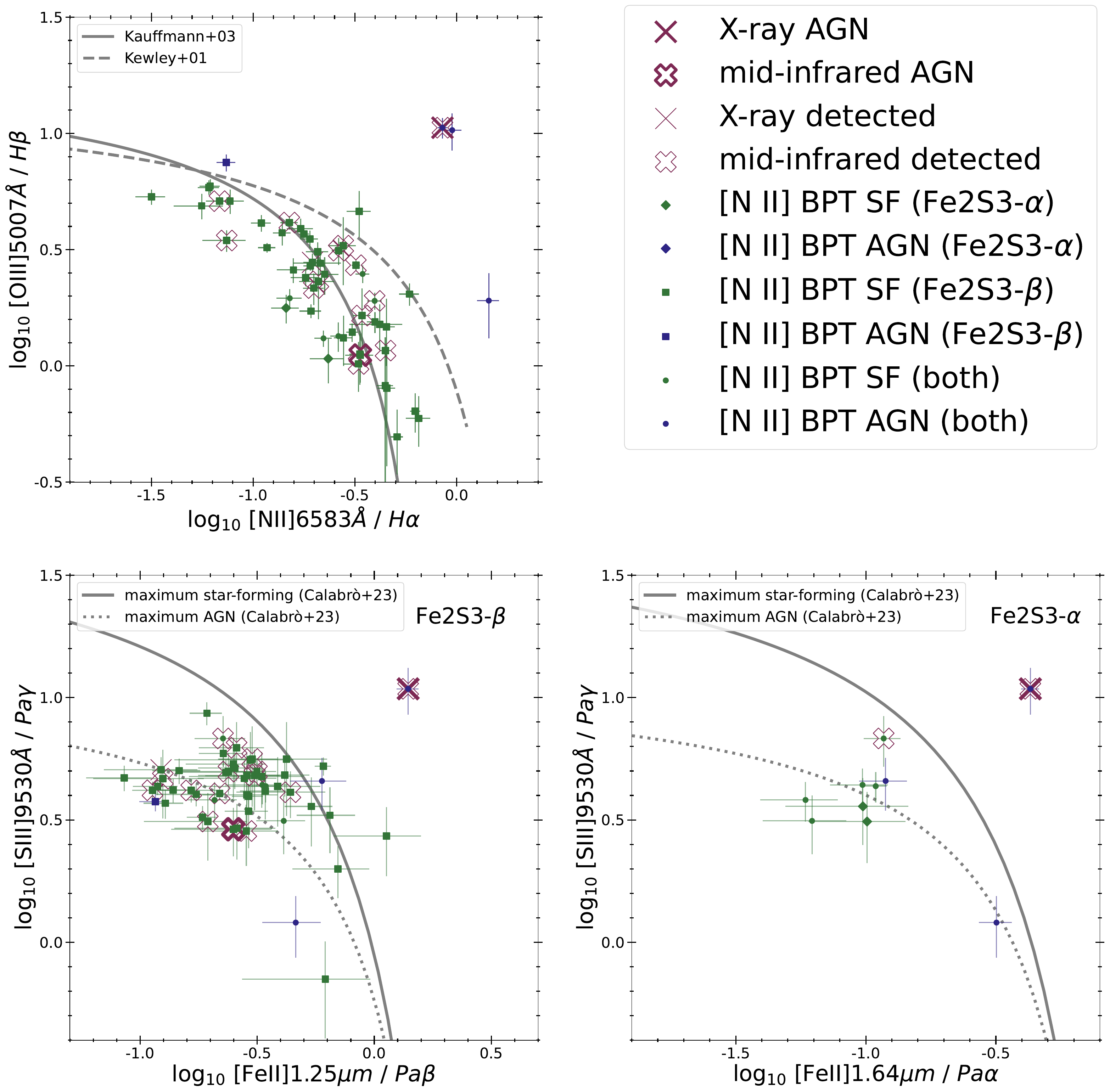}
    \caption{\nii-BPT diagram (upper left) and rest-frame near-IR diagnostics Fe2S3-$\beta$, Fe2S3-$\alpha$ (bottom panels). Optical star-forming galaxies and AGNs are green and blue marks, respectively. Sources in \nii-BPT and Fe2S3-$\alpha$ are diamonds, \nii-BPT and Fe2S3-$\beta$ are squares, and those in all three are circles. X-ray and mid-IR detected sources have extra thin crosses and hollow crosses, and the marks of AGNs are more opaque and thick than not AGNs. In the \nii-BPT diagram, the grey dashed line and solid line are the maximum starburst galaxies line from \citep{Kewley2001t} and the demarcation curve between starburst galaxies and AGNs from \citep{Kauffmann2003}. In the Fe2S3-$\beta$ and Fe2S3-$\alpha$ diagram, the grey solid line and dotted line are the starburst limits and the maximum AGN models from \cite{Calabro2023}.}
    \label{fig: Diagnostic}
\end{figure*}

Figure \ref{fig: Diagnostic} shows the \nii-BPT, Fe2S3-$\beta$, and Fe2S3-$\alpha$ diagrams for 55 galaxies at redshift $1.0<z<3.0$. Within these diagrams, we employ the \cite{Kewley2001} and \cite{Kauffmann2003} (\nii-BPT) and the \cite{Calabro2023} diagnostics (Fe2S3-$\beta$ and Fe2S3-$\alpha$), respectively. We compare the samples present in multiple diagnostics with multiwavelength AGN diagnostics described in Section \ref{sec:data:ancillary}. 49/55 (89.09\%) of sources show agreement between their rest-frame optical and near-IR classifications. 

We note that there are six objects for which the optical and near-IR diagnostics disagree. We discuss the implications of these objects in Section \ref{sec:discussion:individual_sources}. 


\subsection{Dust Attenuation Tracers}\label{sec:results:dust}

\begin{figure*}
    \centering
    \includegraphics[width=1.0\linewidth]{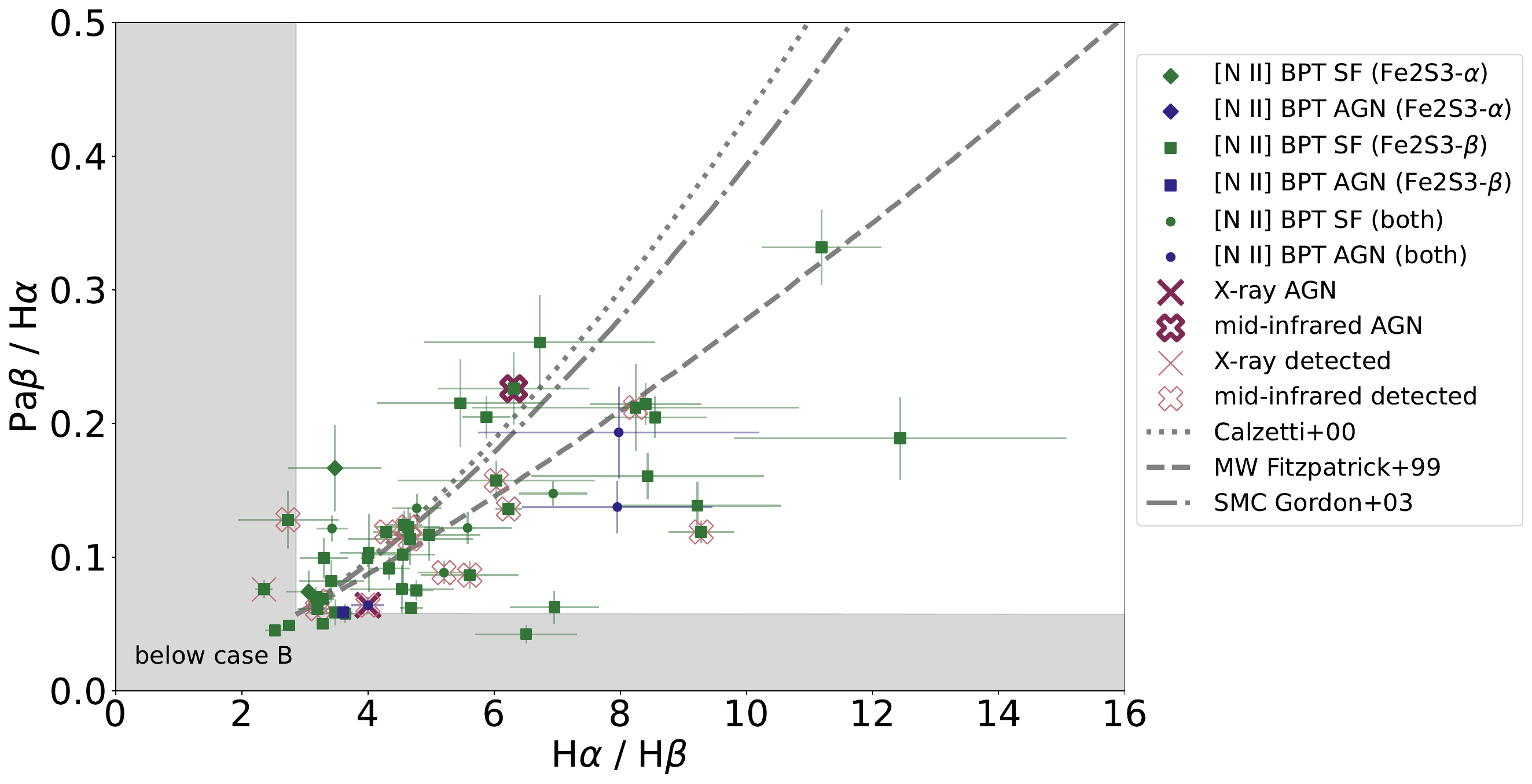}
    \caption{The relation between \Pab \ and the Balmer decrement for the sources involved in near-IR diagnostics. Green and blue marks with different shapes indicate the same categories as the Figure \ref{fig: Diagnostic} . The dotted, dashdot, and dashed grey lines indicate the expected ratios using intrinsic Case B ratios of $\Ha/\Hb = 2.86$ and $\Pab/\Ha = 1/17.6$, and \citep{Calzetti_2000}, \citep{Gordon_2003}, and \citep{Fitzpatrick_1999} attenuation models. The grey shaded region represents line ratios inconsistent with Case B recombination for $n_e = 10^2~\mathrm{cm}^{-3}$ and $T = 10^4$ K \citep[\pab/\ha=1/17.6 and \ha/\hb=2.86;][]{Osterbrock2006}. There are 5 sources located in the grey shaded region, even for 3$\sigma$ uncertainties, can only be accessed with non case B recombination. Most of our sources are consistent with the expectation within 1$\sigma$ uncertainty.}
    \label{fig: dust}
\end{figure*}

In the cases where the optical and near-IR diagnostics disagree, the most straightforward explanation is star-dust and black hole-dust geometry. To test this, we model the dust attenuation curve via multiple ratios of hydrogen emission lines \citep[e.g.,][]{Prescott2022,Cleri2022,Reddy2023}. Figure \ref{fig: dust}  shows the ratios of \Pab/\Ha\ versus the Balmer decrement \ha/\hb. 

There are five sources located at the grey shaded region, which are not consistent with Case B recombination ratios for $n_e = 10^2~\mathrm{cm}^{-3}$ and $T = 10^4$K \citep[\pab/\ha=1/17.6 and \ha/\hb=2.86;][]{Osterbrock2006}. Two sources have significant (>3$\sigma$) offsets from the \citep{Calzetti_2000}, SMC \citep{Gordon_2003} and Milky Way \citep{Fitzpatrick_1999} dust attenuation curves. These sources may be indicative of deviations from Case B recombination, which is an increasingly common interpretation of line ratios at high redshifts \citep[e.g.,][]{Yanagisawa2024,Cameron2024}. We discuss the implications of the discrepant sources in Section \ref{sec:discussion} and summarize the multiwavelength properties of this subsample in Table \ref{tab:sample} of Appendix \ref{appendix:sample}.

\section{Discussion}\label{sec:discussion}

In Figure \ref{fig: Diagnostic}, we have shown the three diagrams of rest-frame optical diagnostic (\nii-BPT) and rest-frame near-IR diagnostics (Fe2S3-$\alpha$, Fe2S3-$\beta$). In this Section, we analyze the results of these diagnostics and discuss the implications of sources whose classifications disagree between the rest-frame optical and near-IR.


The \nii-BPT diagram (55 total sources) classifies 51 galaxies ($93\%$) in our sample as star forming (consistent with the star-forming region below the \cite{Kewley2001} diagnostic within $1\sigma$) and 4 ($7\%$) as AGN above the \cite{Kewley2001} diagnostic within $1\sigma$).

The Fe2S3-$\beta$ diagram, which has 53 sources, classifies 19 galaxies ($36\%$) in our sample as star forming (below the \cite{Calabro2023} maximum AGN diagnostic), 29 galaxies ($55\%$) as composite (between the \cite{Calabro2023} maximum AGN and maximum star-forming diagnostics), and 5 ($9\%$) as AGN (above the \cite{Calabro2023} maximum star-forming diagnostic).

The Fe2S3-$\alpha$ diagram, which has 10 sources, classifies 5 galaxies ($50\%$) in our sample as star forming (below the \cite{Calabro2023} maximum AGN diagnostic), 4 galaxies ($40\%$) as composite (between the \cite{Calabro2023} maximum AGN and maximum star-forming diagnostics), and 1 ($10\%$) as AGN (above the \cite{Calabro2023} maximum star-forming diagnostic).

We find that 49/55 ($89\%$) of sources show agreement between their rest-frame optical and near-IR classifications. This indicates that most sources do not have significant obscured contributions to their ionizing photon budget which arise from different physical processes than their unobscured counterparts.In the following subsections, we interpret the six galaxies for which the rest-frame optical (\nii-BPT) and near-IR (Fe2S3-$\beta$ and/or Fe2S3-$\alpha$) classifications disagree.

\subsection{Discussion of Individual Sources}\label{sec:discussion:individual_sources}
Here we discuss several exemplary objects in our sample, including each of the six sources for which the classifications from rest-frame optical and rest-frame near-IR line ratios disagree. 

\subsubsection{Optical SF, Near-IR AGN Sources}
We identify three sources which are classified as star forming in the rest-frame optical and as AGN in the rest-frame near-IR. 

Of these three galaxies, one has a significant ($>3\sigma$) broad component in \ha\ (j100020.57p021706.45). The elevated Balmer decrement ($\Ha/\Hb = 8.6\pm0.8$) and \Pab/\Ha\ decrement ($\Pab/\Ha = 0.21\pm0.02$), are indicative of high dust attenuation \citep[e.g.,][]{Prescott2022,Cleri2022,Reddy2023}. We interpret this source as a dust-obscured AGN with unobscured star formation, where the emission from the central engine is attenuated heavily enough in the rest-frame optical that it is not classified as an AGN in the \nii-BPT. There are neighboring sources in the field around this object, though the NIRSpec MSA placement excludes the neighbors. 

The other two sources in this category (j033247.42m274711.31 and j100020.04p021557.78) do not have information from multiwavelength AGN tracers. However, both sources have elevated \Pab/\Ha\ and \Ha/\Hb\ ratios, which are indicative of high dust attenuation. We interpret these objects as dusty AGN lacking unobscured sight lines to the broad line region. 

\subsubsection{Optical AGN, Near-IR SF Sources}
We identify three sources that are classified as AGN in the rest-frame optical and as star-forming in the rest-frame near-IR. 

We interpret j141932.60p525023.01 and j142013.00p525533.28 as galaxies with unobscured sightlines to the central AGN with potential dust-obscured star forming regions in the host galaxy. Both of these galaxies exhibit broad Balmer emission (FWHM$_{\Ha} = 1300\pm100$ and $1600\pm1002~\mathrm{km~s^{-1}}$, respectively), indicating that they have unobscured sightlines to the broad line region \citep[e.g.,][]{Antonucci1993,Urry1995}. These sources have high hydrogen line ratios $\Pab/\Ha = 0.14\pm0.02$ and $0.19\pm0.03$, and $\Ha/\Hb = 8.0\pm1.5$ and $8.0\pm2.2$, which are indicative of large amounts of optical dust attenuation.

One other galaxy, j114932.42p222231.19, is not explainable by this model of unobscured AGN combined with obscured star formation. j114932.42p222231.19 has observed $\heii~\lambda4687$, which traces very-high ionization ($>$54 eV) radiation \citep[e.g.,][]{Berg2021,Olivier2022}. Observed emission lines which require ionization energies $>$54 eV are indicative of an ionizing spectrum harder than normal stellar populations, often attributed to accreting black holes or shocks \citep[e.g.,][]{Berg2021,Olivier2022,Izotov2012,Izotov2021,Backhaus2022,Cleri2023a,Cleri2023b,Cleri2025,Chisholm2024,Mingozzi2025}. This source has hydrogen line ratios $\Pab/\Ha = 0.058\pm0.002$ and $\Ha/\Hb = 3.6\pm0.1$, which are consistent with the Case B recombination rates \citep{Osterbrock2006}. One potential explanation for this source is an elevated N/Fe abundance, which would result in a relative increase in \nii/\Ha\ (moving toward the AGN locus in the \nii-BPT) and decrease \feii/\Pab\ and \feii/\Paa\ (moving toward the starforming locus in the Fe2S3-$\beta$ and Fe2S3-$\alpha$). Elevated nitrogen abundances have been observed for many high-redshift galaxies observed with JWST spectroscopy \citep[e.g.,][]{Bunker2023,Larson2023,Castellano2024,Senchyna2024,Flury2025}; nitrogen and iron abundances are directly constrainable with future deep rest-frame optical spectra and/or rest-frame UV spectroscopy for the objects in our sample.

The star-dust and black hole-dust geometries of these galaxies are testable with spatially-resolved spectroscopy from e.g., JWST/NIRSpec integral field unit (IFU). The dusty star forming/AGN contribution would also be testable with mid-IR observations from e.g., the Mid-Infrared Instrument (MIRI) on JWST. Resolved and multiwavelength spectroscopy can provide additional evidence regarding the distribution of the ionizing photon production budget through, e.g., dust-insensitive high-ionization lines \citep[e.g.,][]{Mingozzi2025,Hunt2025} and constraints on dust grain sizes with polycyclic aromatic hydrocarbons \citep[e.g.,][]{Shipley2016,Armus2023}. 

\subsubsection{Other Notable Sources}

One source is identified as an AGN in nearly all diagnostics present in this work (ID: j123607.69p621503.19). In addition to the rest-frame optical and rest-frame near-IR line ratios,  j123607.69p621503.19 qualifies as an X-ray AGN by the \cite{Xue_2011} classification. We also detect a broad component in the \Ha\ line with $S/N = 9.861$, FWHM$=3500\pm200$  km/s, indicating the presence of a broad line region around an accreting supermassive black hole \citep[e.g.,][]{Antonucci1993}. We also find significant (S/N > 3) detections of $\nev~\lambda 3427$ and $\heii~\lambda4687$ emission lines, which require ionization energies of 97.12 eV and 54.42 eV, respectively. The detection of emission lines in this energy regime \citep[the ``very-high'' ionization zone; e.g.,][]{Berg2021} is often considered evidence for an ionizing radiation field harder than that of ``normal'' stellar populations and most commonly attributed to an accreting supermassive black hole \citep[e.g.,][]{Cleri2023a,Cleri2023b,Cleri2025,Olivier2022,Izotov2012,Gilli2010,Negus2023,schmidt1997rosatdeepsurveyii}. The mid-IR for this source is the only tracer which does not find that j123607.69p621503.19 is an AGN, which can be attributed to this object being observed along a sightline which is not heavily dust obscured.

\section{Summary and Conclusions}\label{sec:summary and conclusions}
In this paper, we have compared the JWST/NIRSpec G140M, G235M, and G395M rest-frame optical and near-IR spectroscopy of 55 galaxies at redshifts $1<z<3$ from the DAWN JWST Archive (spectroscopic programs listed in Section \ref{sec:data}). We use the \nii-BPT, Fe2S3-$\beta$, and Fe2S3-$\alpha$ diagnostics to examine their primary sources of ionizing photons, and use multiple hydrogen emission line ratios to probe their attenuation curves. We summarize the main results of this work as follows:
\begin{itemize}
    \item The rest-frame optical \nii-BPT and rest-frame near-IR Fe2S3-$\beta$ and Fe2S3-$\alpha$ line ratio diagnostics agree for 49/55 galaxies in our sample ($89\%$)
    
    \item We identify three sources which are classified as star-forming in the rest-frame optical \nii-BPT, but as AGNs in the rest-frame near-IR Fe2S3-$\beta$. We interpret one of these sources as having a dust-obscured AGN, where the optical spectrum is dominated by star formation, and the other two sources are dusty AGNs lacking unobscured sight lines to the broad line region, or still have the possibility to be star-forming.
    \item We identify three sources which are classified as AGN in the rest-frame optical \nii-BPT and star-forming in the rest-frame near-IR Fe2S3-$\beta$ and Fe2S3-$\alpha$. We interpret two of these sources as having unobscured sightlines to the AGN in the rest-frame optical, but uncovering obscured star formation in the rest-frame near-IR. The third source we interpret as a narrow-line AGN with an elevated N/Fe, which is responsible for the star-forming classification in the rest-frame near-IR.
    \item We compare with multiwavelength catalogs and find that 2/4 and 2/18 sources in our sample are identified as X-ray and mid-IR AGN, respectively.
\end{itemize}

Our results motivate further study of the rest-frame near-IR spectra of galaxies around cosmic noon, particularly in spatially resolved modes (e.g., NIRSpec IFU) to disentangle the contributions of a central engine and the surrounding host galaxy. We demonstrate that the simultaneous use of multiwavelength tracers yields a sample of interesting objects for which observations in the rest-frame optical or rest-frame near-IR alone are not sufficient to fully characterize their sources of ionization. 


\software{Astropy \citep{Astropy2013}, NumPy \cite{Harris2020}, Matplotlib \citep{Hunter2007}}, pandas \citep{Reback2022}, SciPy \citep{Virtanen2020}, LiMe \citep{Fernandez2024}

\begin{acknowledgements}
    NJC also acknowledges support from the Eberly Postdoctoral Fellowship in the Eberly College of Science at The Pennsylvania State University and from NASA grant JWST-AR-05558. 

    The authors greatly appreciate the teams from which we have used publicly available data. Our sample makes use of data from the JWST Advanced Deep Extragalactic Survey (JADES; Program IDs: 1181, 1210, 1286; PIs: D. Eisenstein, N. Leutzgendorf; \citealt{Eisenstein2023a,Eisenstein2023b}), Proposal ID: 1199 (PI: M. Stiavelli; \citealt{Stiavelli2023}), Systematic Mid-Infrared Instrument (MIRI) Legacy Extragalactic Survey (SMILES; Proposal ID: 1207, PI: G. Rieke; \citealt{Rieke2024,Zhu2025}), Cosmic Evolution Early Release Science (CEERS; Proposal ID: 1345, PI: S. Finkelstein; \citealt{Finkelstein2025}), Blue Jay (Program ID: 1810, PI: S. Belli; \citealt{Belli2024}), Assembly of Ultradeep Rest-optical Observations Revealing Astrophysics (AURORA; Program ID: 1914, PI: A. Shapley; \citealt{Shapley2025a,Shapley2025}),  Chemical Evolution Constrained Using Ionized Lines in Interstellar Aurorae (CECILIA; Program ID: 2593, PI: A. Strom; \citealt{Strom2023}), and Red Unknowns: Bright Infrared Extragalactic Survey (RUBIES; Program ID: 4233, PIs: A. de Graaff, G. Brammer; \citealt{deGraaff2025}) surveys. 
    
    Some of the data products presented herein were retrieved from the Dawn JWST Archive (DJA). DJA is an initiative of the Cosmic Dawn Center (DAWN), which is funded by the Danish National Research Foundation under grant DNRF140.

    The Pennsylvania State University campuses are located on the original homelands of the Erie, Haudenosaunee (Seneca, Cayuga, Onondaga, Oneida, Mohawk, and Tuscarora), Lenape (Delaware Nation, Delaware Tribe, Stockbridge-Munsee), Monongahela, Shawnee (Absentee, Eastern, and Oklahoma), Susquehannock, and Wahzhazhe (Osage) Nations. As a land grant institution, we acknowledge and honor the traditional caretakers of these lands and strive to understand and model their responsible stewardship. We also acknowledge the longer history of these lands and our place in that history.
\end{acknowledgements} 

\begin{contribution}
    SRG led the data analysis and preparation of the manuscript. NJC led the project design and contributed to the preparation of the manuscript. JRL provided high-level guidance and support. All other authors provided commentary on the manuscript and assisted in the data analysis and interpretation of results.
\end{contribution}

\begin{appendix}
\section{Sample Characteristics and Auxilliary Data}\label{appendix:sample}
In this Appendix, we present the sample characteristics and multiwavelength AGN diagnostics for each galaxy in our sample present in the Fe2S3-$\alpha$ and Fe2S3-$\beta$ diagrams in Figure \ref{fig: Diagnostic} (Table \ref{tab:sample}).

\begin{deluxetable*}{ccccccccccccc}
\tablecaption{Sample characteristics for sources with disagreement between rest-frame optical and near-IR diagnostics from Figure \ref{fig: Diagnostic} \label{tab:sample}}
\tablehead{
\colhead{DJA ID} & \colhead{Proposal ID} & \colhead{$z_{spec}$}	& \colhead{Fe2S3-$\alpha$\tablenotemark{i}} & \colhead{Fe2S3-$\beta$\tablenotemark{ii}} & \colhead{\nii-BPT\tablenotemark{iii}} & X-ray & Mid-IR	& Rest-Optical	& Rest-NIR & \makecell{\\ Broad (S/N) \\ FWHM [km/s]} & High-ionization & \makecell{\\ \Pab/\Ha \\ \Ha/\Hb} \\
}
\startdata 
j033247.42m274711.31	& 1207 & 1.10	& \dots & $SF-AGN-AGN$	& SF	& \dots & \dots & {\includegraphics[width=0.7in]{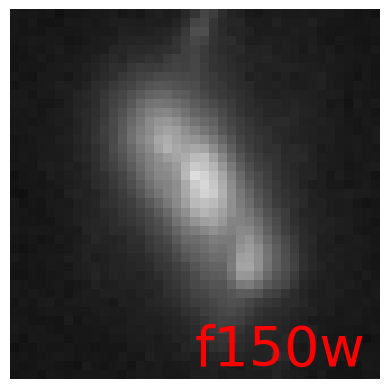}}	& {\includegraphics[width=0.7in]{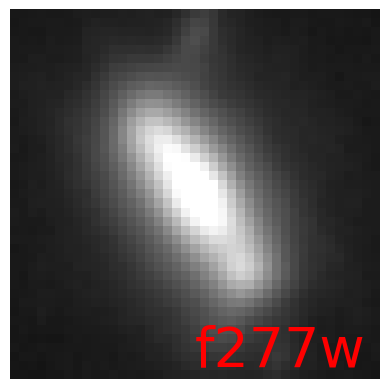}}	& \dots & \dots & \makecell{$0.26\pm0.04$ \\ $6.7\pm1.8$}	\\
j100020.04p021557.78	& 1810 & 1.87	& \dots & $C-AGN-AGN$	& SF	& \dots & \dots & {\includegraphics[width=0.7in]{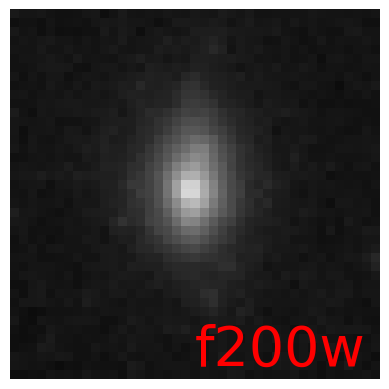}}	& {\includegraphics[width=0.7in]{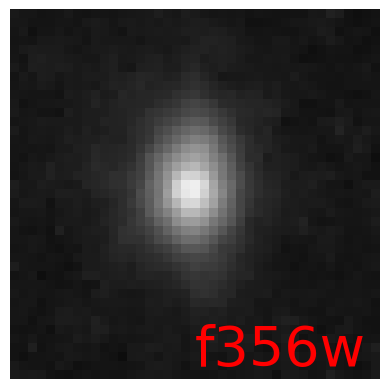}}	& \dots & \dots & \makecell{$0.10\pm0.03$ \\ $4.6\pm0.5$}	\\
j100020.57p021706.45	& 1810 & 2.10	& \dots & AGN	& SF	& \dots & \dots & {\includegraphics[width=0.7in]{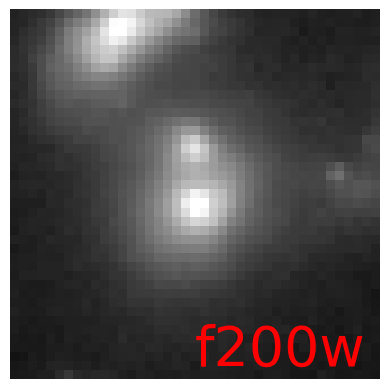}}	& {\includegraphics[width=0.7in]{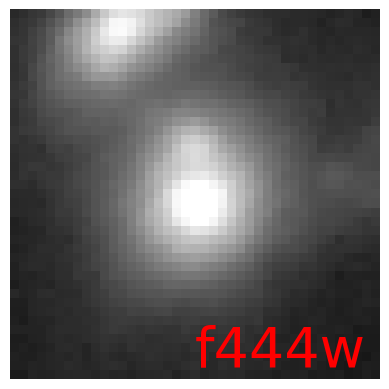}}	& \makecell{\ha\ (13.6) \\ $1600\pm100$}	& \dots & \makecell{$0.21\pm0.02$ \\ $8.6\pm0.8$}	\\
j114932.42p222231.19	& 1199 & 2.76	& \dots & SF	& AGN	& \dots & \dots & {\includegraphics[width=0.7in]{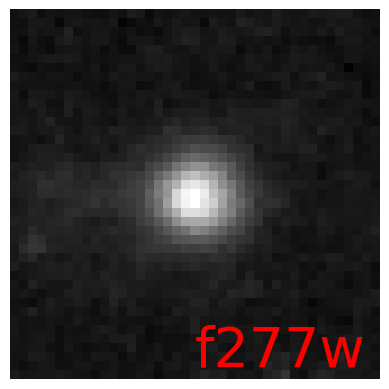}}	& {\includegraphics[width=0.7in]{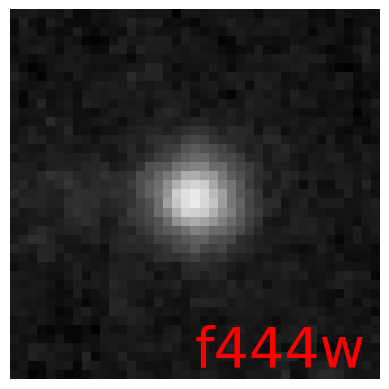}}	& \dots & \heii	& \makecell{$0.058\pm0.002$ \\ $3.6\pm0.1$}	\\
j141932.60p525023.01	& 1345, 4233 & 1.72	& $SF-SF-C$	& $C-AGN-AGN$	& AGN	& \dots & \dots & {\includegraphics[width=0.7in]{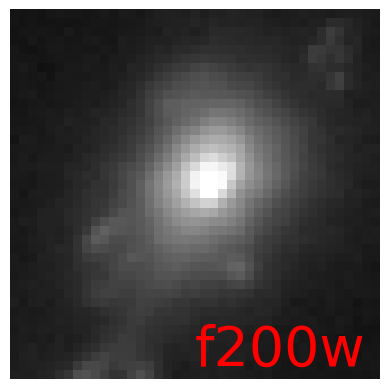}}	& {\includegraphics[width=0.7in]{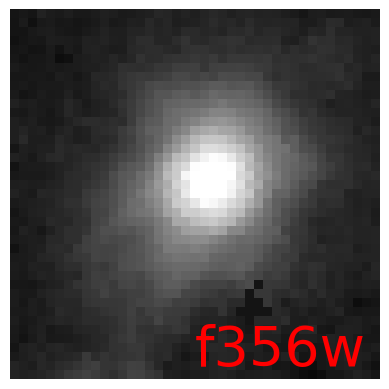}}	& \makecell{\ha\ (5.1) \\ $2200\pm200$}	& \dots & \makecell{$0.14\pm0.02$ \\ $8.0\pm1.5$}	\\
j142013.00p525533.28	& 1345 & 1.04	& SF	& SF	& AGN	& \dots & \dots & {\includegraphics[width=0.7in]{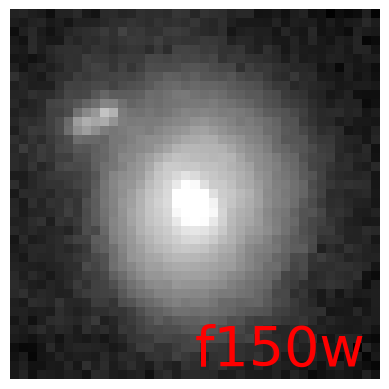}}	& {\includegraphics[width=0.7in]{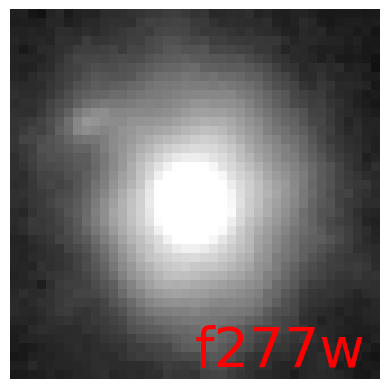}}	& \makecell{\ha\ (10.9) \\ $1600\pm100$}	& \dots & \makecell{$0.19\pm0.03$ \\ $8.0\pm2.2$}	\\
\enddata
\tablenotetext{i}{The notation here shows the \cite{Calabro2023} Fe2S3-$\alpha$ classifications as a tuple for the respective line ratios at ($\mu$ - 1$\sigma$, $\mu$, $\mu$ - 1$\sigma$) where SF indicates that the line ratios lie in the star forming region, C indicates that the line ratios lie in the composite region, and AGN indicates that the line ratios lie in the AGN region. A single notation (e.g., SF) indicates that the line ratios are fully contained in the respective region given the 1$\sigma$ uncertainties.}
\tablenotetext{ii}{The notation here shows the \cite{Calabro2023} Fe2S3-$\beta$ classifications as a tuple for the respective line ratios at ($\mu$ - 1$\sigma$, $\mu$, $\mu$ - 1$\sigma$) where SF indicates that the line ratios lie in the star forming region, C indicates that the line ratios lie in the composite region, and AGN indicates that the line ratios lie in the AGN region.A single notation (e.g., SF) indicates that the line ratios are fully contained in the respective region given the 1$\sigma$ uncertainties.}
\tablenotetext{iii}{The notation here shows the \cite{Kewley2001} \nii-BPT classifications, where SF indicates that the line ratios lie in the star forming region (consistent with the star-forming region below the \cite{Kewley2001} diagnostic within $1\sigma$), and AGN indicates that the line ratios lie in the AGN region ($1\sigma$ above the \cite{Kewley2001}).}
\end{deluxetable*}

\end{appendix}

\clearpage
\bibliography{library}{}
\bibliographystyle{aasjournal}{}

\end{CJK*}
\end{document}